\documentclass[twocolumn]{aa}
\usepackage{graphicx}
\usepackage{epsfig}
\usepackage{rotating}
\begin{document}
   \title{Detection of the H92$\alpha$ recombination line from the starbursts 
          in the Circinus galaxy and NGC 1808}

   \titlerunning{Recombination Line from Circinus Galaxy and NGC 1808}

   \author{A.L. Roy\inst{1,2,3,4}, 
           W.M. Goss\inst{3}, \and
           K.R. Anantharamaiah\inst{4}\fnmsep\thanks{deceased}
          }

   \authorrunning{A. L. Roy et al.}

   \institute{$^{1}$Max-Planck-Institut f\"ur Radioastronomie, Auf dem
              H\"ugel 69, 53121 Bonn, Germany\\
              $^{2}$Australia Telescope National Facility, PO Box 76,
              Epping 1710, NSW, Australia\\
              $^{3}$NRAO, PO Box O, Socorro, NM 87801, USA\\
              $^{4}$Raman Research Institute, CV Raman Ave,
              Sadashivanagar, Bangalore 560080, India}
   \date{Received ; accepted}

   \abstract{
     
     {\it Context.}  Gas ionized by starburst activity radiates radio
     recombination lines (RRLs), from which one can derive the
     plasma conditions and the number of massive stars formed in the burst, free
     of dust obscuration effects.
     
     {\it Aims.} We aimed to find detectable RRL emission from additional
     extragalactic starburst systems and to use the line properties to estimate
     the properties of the ionized gas.
     
     {\it Methods.} We conducted a search for RRLs in the nearby extragalactic
     starburst or Seyfert galaxies NGC\,1808, the Circinus galaxy,
     NGC\,4038/9, II\,Zw\,40, NGC\,6221, NGC\,7552, IRAS\,18325-5926, IC\,5063,
     and VV\,114. We used the Very Large Array with resolution of $3''$
     to $32''$ and the Australia Telescope Compact Array with
     resolution of $10''$ to search for the RRLs H91$\alpha$
     and H92$\alpha$ with rest frequencies of 8.6 GHz and 8.3 GHz.
     From the new detections we derive conditions in the starburst regions.

     {\it Results.} We detected for the first time RRLs from the starburst
     nuclei in the Circinus galaxy and NGC\,1808.  The Circinus galaxy was
     detected in RRL emission with a line strength integrated over the
     source of $ 3.2$\,mJy, making it the fourth-strongest
     extragalactic RRL emitter known at this frequency (after NGC\,4945, M\,82,
     and NGC\,253) and so is suitable for detailed study.  The line and
     continuum emission from the Circinus galaxy can be matched by a model
     consisting of a collection of 50 to 10\,000 H~II regions with
     temperatures of 5000\,K, densities of 500\,cm$^{-3}$ to 50\,000\,cm$^{-3}$,
     and a total effective diameter of 3\,pc to 50\,pc.  The Lyman continuum production rate
     required to maintain the ionization is $1 \times 10^{52}~\mathrm{s}^{-1}$
     to $3 \times 10^{53}~\mathrm{s}^{-1}$, which requires 300 to 9000 O5
     stars to be produced in the starburst, inferring a star formation rate of
     0.2\,$M_\odot$ yr$^{-1}$ to 6\,$M_\odot$ yr$^{-1}$.  NGC\,1808 was detected in RRL emission
     at $3.9\sigma$ with a line strength of 0.47\,mJy.
     No radio recombination lines were detected from
     the other galaxies surveyed to a $3\sigma$ limit of 0.3\,mJy to 1.4\,mJy.
     
     {\it Conclusion.} We have detected RRLs from two galaxies, adding to the small but growing
     number of known extragalactic RRL emitters.  The Circinus galaxy
     is strong and especially suited to high-quality follow-up spectroscopic
     study.  We derived conditions and star formation rates
     in the starbursting regions.  Uncertainties can be reduced by future
     multi-transition studies.

\keywords{galaxies: individual: NGC 1808 - galaxies: individual: Circinus
 Galaxy - galaxies: nuclei - radio lines: galaxies}
   }

   \maketitle
%

\section{Introduction}

Starburst activity has a profound impact on galaxies, through prodigious
formation of stars and clusters, ionization of gas and outflows driven by
overpressure in the starbursting region that redistribute the interstellar
medium (ISM).  Studies of the gas ionized in the burst are interesting because
they provide information on the ionizing photon production rate and hence the
number of high-mass stars formed and on the conditions in the surrounding ISM.

However, optical and infrared diagnostics of the ISM are hampered by dust
obscuration associated with the molecular clouds from which the
stars form.  Radio studies are free of dust obscuration effects
and the detection of radio recombination lines is particularly
useful as their strengths provide diagnostics of the plasma 
conditions and give dynamical information with arcsec
resolution.

RRLs were first detected from the starbursts in M\,82 and NGC\,253 by Shaver et
al. (1977) and Seaquist \& Bell (1977) soon after the potential to detect and
interpret extragalactic RRLs was shown by Shaver (1978).  Those galaxies have
since been studied over a wide range of frequencies, giving constraints on the
physical state and kinematics in the nuclear regions (e.g. Anantharamaiah \&
Goss 1997; Rodr\'iguez-Rico et al. 2006).

However, surveys to find RRLs in other galaxies produced no further detections
for a period (Churchwell \& Shaver 1979; Bell \& Seaquist 1978; Bell et al.
1984).  When survey sensitivities were improved by an order of magnitude using
the Very Large Array (VLA) during a renewed effort in the early 1990s, RRLs
were detected near 8.6\,GHz from several bright starburst galaxies.  Those new
detections are NGC\,660, NGC\,1365, NGC\,2146, NGC\,3628, NGC\,3690,
NGC\,5253, M\,83, IC\,694, Arp\,220, Henize\,2-10 (Anantharamaiah et al. 1993;
Zhao et al. 1996; Phookun et al.  1998; Mohan et al. 2001), and NGC\,4945 at mm
wavelengths (Viallefond, private communication).

During an extension of the RRL surveys to southerly starburst and Seyfert
galaxies using the Australia Telescope Compact Array (ATCA), we made three new
detections: NGC\,3256, the Circinus galaxy, and NGC\,4945.  The detection of
NGC\,3256 was reported in a previous paper (Roy et al. 2005a). Here, we report
the detection of the Circinus galaxy and four upper limits (NGC\,6221,
NGC\,7552, IC\,5063, and IRAS\,18325-5926).  The detection of
NGC\,4945 will be reported in a later paper.  A short report of all
three detections was published by Roy et al. (2005b).

During an extension of the RRL surveys using the VLA, we made a new
detection of NGC\,1808 (Mohan 2002) and established sensitive
upper limits on four other galaxies (NGC\,4038/9, NGC\,7552, II\,Zw\,40, and
VV\,114), all of which are reported here.

We adopt $H_{0} = 75~\mathrm{km~s^{-1} Mpc^{-1}}$, $q_{0} = 0.5$, and 
$\Lambda = 0$, and give velocities in the heliocentric frame using
the optical velocity definition throughout this paper.


\section{Observations} 

\subsection{ATCA Observations} 

We observed the Circinus galaxy, IC\,5063, IRAS\,18325-5926,
NGC\,7552, and NGC\,6221 with the Australia
Telescope Compact Array (ATCA).  The observing parameters and results
are summarized in Table 1.

Calibration and imaging were done using the AIPS software using standard
methods.  The flux-density scale assumed that PKS B1934-638 had a flux density
of 2.99\,~Jy at 8295\,MHz and 2.87\,Jy at 8570\,MHz, based on the Baars et al.
(1977) flux-density scale.  A phase calibrator was observed every half hour to
correct the instrumental phase response.  A bandpass calibrator was observed
every few hours for correcting the instrumental frequency response (bandpass).
Phase corrections obtained from self calibration of the continuum source were
applied to the spectral line data.  Continuum emission was subtracted from the
line data using the method UVLSF (Cornwell, Uson \& Haddad 1992) in which the
continuum is determined for each baseline by a linear fit to the visibility
spectrum.  The final continuum and line images were made using natural
weighting of the $(u, v)$ data to achieve maximum possible signal-to-noise
ratio and we averaged together the two transitions and two polarizations.

Uncertainties on the absolute flux densities have an 11\,\% rms random
multiplicative component due to flux-density bootstrapping and atmospheric
opacity, a 0.12\,mJy rms random additive component due to thermal noise in a
1\,MHz channel and a systematic multiplicative component of 11\,\% rms due
mainly to the uncertainty in the Baars et al.  flux-density scale.

\subsection{VLA Observations}

NGC 1808, NGC\,4038/9, II\,Zw\,40, NGC\,7552, and VV\,114 were observed with
the Very Large Array (VLA).  The observing parameters and results are
summarized in Table 2.  Data reduction was similar to the ATCA reduction.

For NGC\,1808, the CnB data suffered from a high residual bandpass calibration
error causing low SNR spectra.  Thus, we based the analysis on the DnC spectral
data, but show the continuum image from the higher resolution CnB data.

\onecolumn

\begin{sidewaystable}
\caption[]{ATCA Observational Parameters and Results.}
\label{ObsLine}
\scriptsize
\begin{center}
\begin{tabular}{ll|l|l|l|l}
\hline
\hline
\noalign{\smallskip}
                            & \multicolumn{1}{c}{Circinus galaxy}
                            & \multicolumn{1}{c}{NGC 6221}
                            & \multicolumn{1}{c}{NGC 7552}
                            & \multicolumn{1}{c}{IC 5063}
                            & \multicolumn{1}{c}{IRAS 18325-5926} \\
\noalign{\smallskip}
\hline
\noalign{\smallskip}

Array configuration         & 750D (1993jul),750C (1994oct)
                            & 750C  
                            & 750C  
                            & 750A  
                            & 750A \\

                            & 375 (1994nov)
                            &
                            &
                            &
                            & \\

Date of observation         & 1993jul, 1994oct20 \& nov26 
                            & 1994oct22, 23
                            & 1994oct22, 23
                            & 1994mar3, 4, 5, 6, 7
                            & 1994mar3, 4, 5, 6, 7 \\

Integration time (h)        & 37
                            & 8.5
                            & 6.2
                            & 9.7
                            & 10.6 \\

Transitions                 & H92$\alpha$ \& H91$\alpha$  
                            & H92$\alpha$ \& H91$\alpha$ 
                            & H92$\alpha$ \& H91$\alpha$ 
                            & H92$\alpha$ \& H91$\alpha$  
                            & H92$\alpha$ \& H91$\alpha$ \\

$\nu_{\rm rest}$ of RRL (MHz)
                            & 8309.38 \& 8584.82
                            & 8309.38 \& 8584.82
                            & 8309.38 \& 8584.82
                            & 8309.38 \& 8584.82
                            & 8309.38 \& 8584.82 \\

$\nu_{\rm{band-centre}}$ (MHz)
                            & 8295.00 \& 8570.00
                            & 8265.00 \& 8540.00
                            & 8265.00 \& 8540.00
                            & 8225.00 \& 8487.00
                            & 8141.00 \& 8411.00 \\

$V_{\mathrm{systemic,helio,optical}}$ (km\,s$^{-1}$) 
                            & 480
                            & 1477
                            & 1590
                            & 3402
                            & 6065 \\

Distance (Mpc)              & 4.2
                            & 18
                            & 21
                            & 45
                            & 81 \\

Bandwidth (MHz), channels, IFs
                            &  64, 64, 4
                            &  64, 64, 4
                            &  64, 64, 4
                            &  64, 64, 4
                            &  64, 64, 4 \\

Spectral resolution (km\,s$^{-1}$)
                            &  36
                            &  36
                            &  36
                            &  36
                            &  37 \\

Polarization                & dual linear
                            & dual linear
                            & dual linear
                            & dual linear
                            & dual linear \\

Minimum baseline (k$\lambda$)  & 0.86
                              & 1.27
                              & 1.27
                              & 2.13
                              & 2.13 \\

Beam size (natural weight) ($'' \times ''$)
                            &  $11.0'' \times 9.4'' $
                            &  $11.8'' \times 7.9'' $
                            &  $14.0'' \times 8.3'' $
                            &   $6.4'' \times 5.8'' $
                            &   $6.3'' \times 5.6'' $ \\

Beam position angle         & $-76^{\circ}$
                            & $32^{\circ}$
                            & $16^{\circ}$ 
                            & $24^{\circ}$ 
                            & $50^{\circ}$ \\

Phase calibrator (B1950)    & 1414-596
                            & 1740-517
                            & 2313-438
                            & 1934-638
                            & 1934-638 \\

Bandpass calibrator (B1950) & 0537-441, 1921-293
                            & 1921-293 
                            & 0316+413, 0537-441, 
                            & 0537-441, 2251+158
                            & 2251+158 \\

                            &
                            &
                            & 1921-293
                            &
                            & \\

\bf{Line and Continuum Properties} & & & & & \\

Peak continuum surface brightness (mJy\,beam$^{-1}$) & $ 98.3 \pm 15 $ 
                                    & $ 20.2 \pm 2.3 $
                                    & $ 44.7 \pm 5.0 $ 
                                    & $ 180 \pm 20 $
                                    & $ 30.9 \pm 3.5 $ \\

Total continuum flux density (mJy) & $ 234 \pm 37 $
                              & $  20 \pm 3 $ 
                              & $  57 \pm 6 $ 
                              & $ 177 \pm 20 $ 
                              & $  30 \pm 4 $ \\

Noise per image channel (mJy\,beam$^{-1}$)
                            & 0.12
                            & 0.35
                            & 0.45
                            & 0.33
                            & 0.27 \\

Integrated line flux density $^f$ (mJy)      & $ 3.2 \pm 0.6 $  
                                             & $< 1.0 $
                                             & $< 1.4 $
                                             & $< 1.0 $
                                             & $< 0.8 $ \\
      
Integrated line flux or 3$\sigma$ upper limit $^g$ (W\,m$^{-2}$) & $(20.1 \pm 3.1) \times 10^{-23} $
                                             & $ < 1.0 \times 10^{-23} $
                                             & $ < 1.4 \times 10^{-23} $
                                             & $ < 1.0 \times 10^{-23} $
                                             & $ < 0.8 \times 10^{-23} $ \\

Measured line width (FWHM; km\,s$^{-1}$)    & $ 278 \pm 10 $   
                                             & --
                                             & --
                                             & --
                                             & -- \\

Deconvolved line width (FWHM; km\,s$^{-1}$) & $ 260 \pm 10 $   
                                             & --
                                             & --
                                             & --
                                             & -- \\

Centroid helio. optical vel. (km\,s$^{-1}$)  & $464 \pm 10 $
                                             & --
                                             & -- 
                                             & --
                                             & -- \\

Cont. flux density over line region (mJy)    & $225 \pm 35 $ 
                                             & -- 
                                             & --
                                             & --
                                             & -- \\

No. of beam areas where line is observed     & 1.9 
                                             & -- 
                                             & -- 
                                             & --
                                             & -- \\

Total 2.7 GHz continuum (mJy)
                               & 
                               & 220$^{e}$
                               & 220$^{a}$
                               & 800$^{a}$
                               & 115$^{c}$ (at 2.4 GHz)\\

Total 5.0 GHz continuum (mJy)
                               & 
                               & 110$^{d}$
                               & 88$^{b}$ (at 4.8 GHz)
                               & 420$^{a}$
                               &  \\

\noalign{\smallskip}
\hline
\end{tabular}
\end{center}

$^{a}$ Wright \& Otrupcek (1990) \\
$^{b}$ Forbes et al. (1994b) \\
$^{c}$ Roy (1995) \\
$^{d}$ Whiteoak (1970) \\
$^{e}$ Wright (1974) \\
$^{f}$ Upper limits are three times the rms noise in a single channel. \\
$^{g}$ Upper limits assume a FWHM equal to the velocity resolution of the data. \\

\end{sidewaystable}

\begin{sidewaystable}
\scriptsize
\begin{center}
\caption[]{VLA Observational Parameters and Results}
\vspace{0.2cm}
\begin{tabular}{lcc|cc|c|cc|c}
\hline
\hline
\noalign{\smallskip}
                     & \multicolumn{2}{c}{NGC 1808}
                     & \multicolumn{2}{c}{NGC 4038/9} 
                     & \multicolumn{1}{c}{II Zw 40}
                     & \multicolumn{2}{c}{NGC 7552} 
                     & VV 114 \\

\noalign{\smallskip}
\hline
\noalign{\smallskip}

Array configuration  & CnB
                     & DnC
                     & C 
                     & DnC 
                     & D
                     & CnB 
                     & D 
                     & D \\

Date of observation  & 1998nov10
                     & 1999feb18
                     & 1998dec27
                     & 1999feb20
                     & 2000aug14, sep29
                     & 1998nov16
                     & 1999apr23
                     & 1999apr03 \\

Integration time (h)        & 6.3
                            & 6.3
                            & 
                            & 
                            &
                            &
                            &
                            &  \\

Transitions                 & H92$\alpha$  
                            & H92$\alpha$  
                            & H92$\alpha$  
                            & H92$\alpha$ 
                            & H92$\alpha$ \& H93$\alpha$
                            & H92$\alpha$  
                            & H91$\alpha$ 
                            & H92$\alpha$ \\

$\nu_{\rm rest}$ of RRL (MHz)
                            & 8309.38
                            & 8309.38
                            & 8309.38
                            & 8309.38
                            & 8309.38, 8045.60
                            & 8309.38
                            & 8584.82
                            & 8309.38 \\

$\nu_{\rm{band-centre}}$ (MHz)
                            & 8285.100
                            & 8285.100
                            & 8264.725
                            & 8264.725
                            & 8288.185, 8025.080
                            & 8264.900
                            & 8535.100
                            & 8288.185 \\

$V_{\mathrm{systemic,helio,optical}}$ (km\,s$^{-1}$) 
                             & \multicolumn{2}{c|}{1014}
                             & \multicolumn{2}{c|}{1663}
                             & 789
                             & \multicolumn{2}{c|}{1653}
                             &  6016\\

Distance (Mpc)                 & \multicolumn{2}{c|}{11}
                               & \multicolumn{2}{c|}{19.2} 
                               & 10.5 
                               & \multicolumn{2}{c|}{21}
                               & 77 \\

Bandwidth (MHz), channels, IF  & 46.9, 15, 1
                               & 46.9, 15, 1
                               & 23.44, 15, 2 
                               & 23.44, 15, 2 
                               & 23.4, 15, 2 
                               & 46.9, 15, 1 
                               & 46.9, 15, 1 
                               & 23.4, 15, 2 \\

Spectral resolution$^a$ (km\,s$^{-1}$)
                               & 226
                               & 226
                               & 113.4
                               & 113.4
                               & 113
                               & 226
                               & 226
                               & 115 \\

Polarization                & dual circular
                            & dual circular
                            & dual circular
                            & dual circular
                            & dual circular
                            & dual circular
                            & dual circular
                            & dual circular \\

Minimum baseline (k$\lambda$)  & 3.8
                               & 1.2
                               & 2.0
                               & 1.3
                               & 0.95
                               & 2.6
                               & 2.5
                               & 0.8 \\

Beamsize (natural weight)
                               & $3.6''\times 2.5''$  
                               & $8.2''\times 6.6''$ 
                               & $4.8''\times 2.7''$  
                               & $12.3''\times 7.2''$ 
                               & $11.9''\times 11.7''$
                               & $5.1''\times 2.5''$  
                               & $31.4''\times 6.7''$ 
                               & $16.0''\times 10.0''$ \\

Beam position angle            & $-6^{\circ}$
                               & $1^{\circ}$
                               & $7^{\circ}$
                               & $8^{\circ}$
                               & $-11^{\circ}$
                               & $11^{\circ}$
                               & $-3^{\circ}$
                               & $18^{\circ}$ \\

Phase calibrator (B1950)       & 0402-362 
                               & 0402-362
                               & 1127-145 
                               & 1127-145 
                               & 0530+135 
                               & 2311-452 
                               & 2311-452 
                               & 0113-118 \\

                               &
                               &
                               &  
                               &  
                               &  
                               & \& 2227-394 
                               & 
                               & \\

Bandpass calibrator (B1950)    & 0402-362
                               & 0402-362
                               & 1127-145 
                               & 1127-145 
                               & 0530+135 
                               & 2255-282 
                               & 2255-282 
                               & 0113-118 \\

                               &
                               &
                               &  
                               &  
                               & \& 3C 48 
                               & \& 3C 48 
                               &  
                               & \\

\bf{Line and Continuum Properties} & & & & & & & & \\

Peak continuum surface brightess (mJy\,beam$^{-1}$)
                               & 19.6
                               & 45.8
                               & 5.08
                               & 9.13
                               & 13.4
                               & 13.6$^b$
                               & 40.2
                               & 27.3 \\

Total continuum flux density (mJy)
                               & 110
                               & 124
                               & 123
                               & 128
                               & 20
                               & 55
                               & 55
                               & 54 \\

Noise in continuum image ($\mu$Jy\,beam$^{-1}$) 
                               & 35
                               & 33
                               & 25 
                               & 22 
                               & 20 
                               & 33 
                               & 45 
                               & 23 \\

Noise per image channel ($\mu$Jy\,beam$^{-1}$) 
                               & 120
                               & 140
                               & 66 
                               & 68 
                               & 56 
                               & 80 
                               & 135
                               & 56 \\

\cline{2-3}

Peak line surface brightness (mJy\,beam$^{-1}$)     & $< 0.36$
                                              & $0.47 \pm 0.08$
                                              & \multicolumn{2}{c|}{$< 0.20$}
                                              & $< 0.17$
                                              & \multicolumn{2}{c|}{$< 0.24$}
                                              & $< 0.17$ \\

Integrated line flux or 3$\sigma$ to 6$\sigma$ upper limit $^c$ (W\,m$^{-2}$) 
               & $< 4 \times 10^{-23} $
               & $(4.5 \pm 1) \times 10^{-23} $
               & \multicolumn{2}{c|}{$< 0.47 \times 10^{-23}$} 
               & $< 0.6 \times 10^{-23}$
               & \multicolumn{2}{c|}{$< 0.76 \times 10^{-23}$} 
               & $< 0.57 \times 10^{-23}$ \\

Measured line width (FWHM; km\,s$^{-1}$)      &  
                                              & 339 $\pm$ 75  \\

Deconvolved line width (FWHM; km\,s$^{-1}$)   &   
                                              & 215 $\pm$ 50  \\

Peak 8.3 GHz continuum (mJy\,beam$^{-1}$)       & 7.0 &  \\

\noalign{\smallskip}
\hline
\end{tabular}
\end{center}
$^{a}${After offline Hanning smoothing.}\\
$^{b}${The peak continuum strength for sources A \& B is 
11.5 mJy\,beam$^{-1}$ and for C \& D is 13.6 mJy\,beam$^{-1}$ in the CnB image.}\\
$^{c}${Assuming a FWHM equal to the velocity resolution of the data.}\\
\end{sidewaystable}

\clearpage

\twocolumn

\section{The Circinus Galaxy}

This remarkable low-latitude spiral is at a distance of about 4\,Mpc (thus $1''$ is
19\,pc) and is the closest Seyfert galaxy (type 1) with many
signs of a starburst in the nucleus.

The H\,I distribution has been studied with the ATCA (Jones et al. 1999) who
find a total size of about $80'$ or 90\,kpc with a total H\,I mass of $7 \times 10^{9}
M_{\odot}$.  Detailed imaging of the circumstellar ring with a radius of
about 220\,pc in H$\alpha$ (Elmouttie et al. 1998a) shows a rotational speed
of about 350\,km\,s$^{-1}$ in rough agreement with an H\,I feature (Jones et 
al. 1999).  This feature may lie close to the inner Lindblad resonance.

A number of studies of the radio continuum have been carried out (Elmouttie et
al. 1998b, Harnett et al. 1990, Elmouttie et al. 1995). The peculiar radio
lobes orthogonal to the disk of the galaxy are polarized and the nuclear
source (about 0.1 Jy at 1.4 GHz) has a flat spectral index of $\alpha = -0.06$
($S \propto \nu^{\alpha}$) with a size limit of $\sim 20$\,pc.  Higher
resolution radio images by Davies et al. (1998) at 8.4\,GHz and 5\,GHz with a
resolution of $1''$ to $2''$ show a compact source with a size of about
$0.6''$ or 12\,pc. There is also a reported radio core of 19\,mJy at 2.3\,GHz
observed with a beam of $0.1''$ with the Parkes-Tidbinbilla Interferometer,
which can be associated with the AGN. Davies et al. suggest that the flat
spectral index of the nucleus could arise from free-free absorption.

There have been several studies of the molecular content of the Circinus
galaxy by Curran et al. (2001) and Elmouttie et al. (1998c). The former
authors have carried out a systematic study of numerous molecules (eg four
isotopes of CO in several transitions, CS, H$_{2}$CO, HCN, HCO$^{+}$ and other
molecules).  They propose that the H$_{2}$ density near the nucleus is in the
range $2 \times 10^{3}$\,cm$^{-3}$ to $10^{5}$\,cm$^{-3}$. The molecular
emission is distributed in a disk of radius 300\,pc which is the likely source
of the gas for the star formation.  H$_{2}$O masers have been extensively
studied by Greenhill et al.  (2003a, 2003b).  Doppler shifts up to $\sim
460$\,km\,s$^{-1}$ are observed.

In the optical and near IR, high resolution investigations of the Circinus
galaxy have been carried out by Wilson et al. (2000) who have used HST imaging in
[O\,III], H$\alpha$, H$_{2}$ and continuum bands in the optical and the near
IR.  A one-sided ionization cone is observed in H$\alpha$ and [O\,III] and
the nuclear starburst activity is shown in H$\alpha$ images at distances of
tens of parsecs from the centre. The circumstellar star-forming ring of radius
$\sim 200$\,pc is prominent.

\subsection{Results}

The ATCA continuum and line images, integrated spectrum, and
position-velocity diagram are shown in Figs. 1 to 3.  The measured line
and continuum parameters are given in Table 1. 

The continuum image shows a single extended continuum component.  The
continuum emission is predominantly non-thermal, with a spectral index of
-0.65 between 1.4\,GHz and 8.3\,GHz (Elmouttie et al. 1998b).

Line emission was detected in the nuclear region, coincident with the
peak continuum emission with a tail of emission extending $25''$ (500 pc) 
towards the north-east.  The total area of line emission is 1.9 beam
areas, or (250\,pc)$^2$.

The H91$\alpha$ + H92$\alpha$ spectrum integrated over the line-emitting
region shows a clear line detection with $26\,\sigma$ significance with
centroid at 480\,km\,s$^{-1}$, compared to the H\,I systemic velocity of
449\,km\,s$^{-1}$ (Juraszek et al. 2000).  The line FWHM is 260\,km\,s$^{-1}$
after deconvolving the instrumental velocity resolution of 42\,km\,s$^{-1}$.

The position-velocity diagram (Fig 3) shows a gradient in the east-west
direction of 130\,km\,s$^{-1}$ over the central $25''$ (475\,pc)
consistent with a disk rotating about a north-south axis with the east side
receding.  The position angle for the position-velocity slice was chosen
by inspecting the first-moment image, which showed the velocity gradient to
be east-west.  Interestingly, the gradient direction is opposite that of
the Br$\gamma$ or H$_{2}$ (M\"uller Sanchez et al. 2006), the H$_{2}$O masers
(Greenhill et al.  2003) and the H\,I disk (Jones et al. 1999), all of which
show the south-west side receding.  The gradient direction of the RRL emission
is the same as that of the molecular outflow seen by Elmouttie et al. (1997)
and studied by Curran et al. (1999).  Thus, the H92$\alpha$ emitting gas is
most likely associated with the outflow.

The velocity span of the signal in the position-velocity diagram 
(130\,km\,s$^{-1}$) is half as wide as
the line integrated over a $38'' \times 38''$ region in Fig 2
(260\,km\,s$^{-1}$).  This indicates that there is extended low-level RRL
emission outside the region in the position-velocity diagram, i.e. at
higher or lower declination.  The line width integrated
over the large area (260\,km\,s$^{-1}$) is comparable to that of the molecular
outflow ($\pm 190$\,km\,s$^{-1}$) and of the ionized outflow
(180\,km\,s$^{-1}$; Veilleux \& Bland-Hawthorn 1997) and
is much larger than the width of the Br$\gamma$ and H$_{2}$ emission
(30\,km\,s$^{-1}$) seen on 8\,pc scale (M\"uller Sanchez et al. 2006), further
supporting the association of the RRL-emitting gas with the outflow.

\begin{figure}
\centering
\includegraphics[width=8cm]{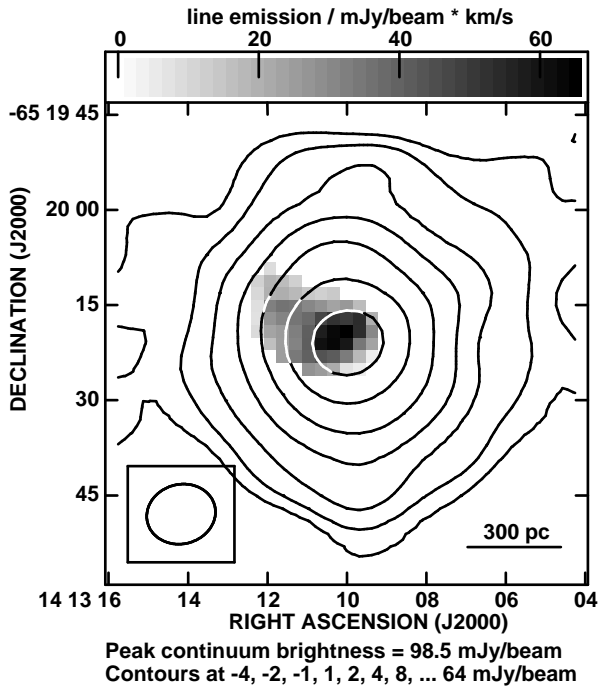}
\caption{
ATCA 8.3 GHz + 8.6 GHz continuum image of the Circinus galaxy (contours)
superimposed on the grey scale zeroth-moment image showing H91$\alpha$ +
H92$\alpha$ line emission.  Peak continuum brightness is
98.5\,mJy\,beam$^{-1}$ and contours are at -4,-2,-1,1,2,4,8,16,32, and 
64\,mJy\,beam$^{-1}$.  Grey-scale peak = 65\,mJy\,beam$^{-1}$\,km\,s$^{-1}$.
Beamsize is 
$11.0'' \times 9.4''$ at a P.A. of -76$^{\circ}$, rms noise 
is 0.12\,mJy\,beam$^{-1}$\,channel$^{-1}$.
}
\label{CircinusContLine}
\end{figure}

\begin{figure}
\centering
\includegraphics[width=8cm]{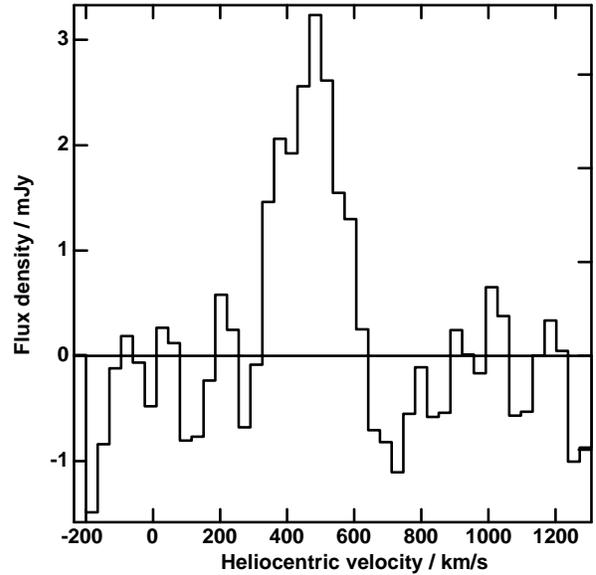}
\caption{
ATCA H91$\alpha$ + H92$\alpha$ line profile integrated over the
line-emitting region in the Circinus galaxy.
Region of integration is a box of size $38'' \times 38''$ centred
on RA 14 13 10.01  dec -65 20 19.1.
}
\label{CircinusSpec}
\end{figure}

\begin{figure}
\centering
\includegraphics[width=8cm]{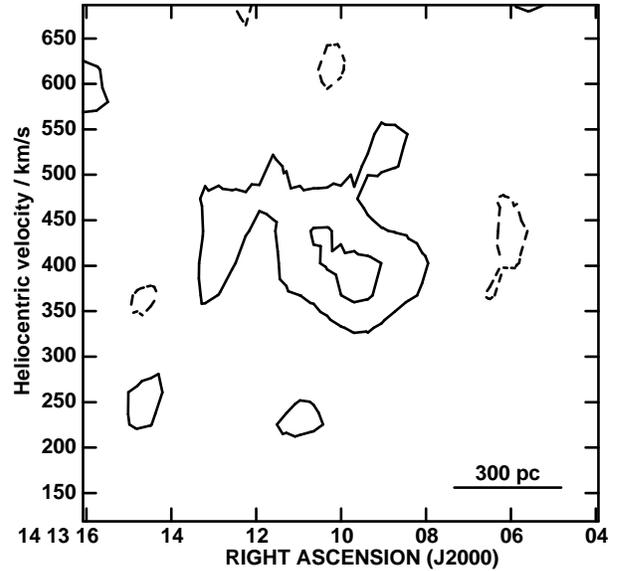}
\caption{
ATCA position-velocity diagram of H91$\alpha$ + H92$\alpha$ emission in the Circinus
galaxy along a slice at constant declination (-65$^{\circ}$\,20'\,19.1'') through the
maximum of the RRL emission, showing possible rotation.  
Beamsize is $11.0''\times9.4''$ at a P.A. of -76$^{\circ}$, giving 10.6'' resolution
in RA, 36\,km\,s$^{-1}$ in velocity (no Hanning smoothing), and averaging over 9.8'' 
in declination.
}
\label{CircinusMOM1}
\end{figure}

\subsection{Modelling the Ionized Gas}

Two types of models have been considered for the RRL emission from
the nuclei of external galaxies: one based on a uniform slab of ionized gas 
and the other based on a collection of compact H~II regions.  Such models
have been discussed by Anantharamaiah et al. (2000) and references therein,
and are documented in detail by Mohan (2002).
These models take as constraints the integrated RRL strength at one or more
frequencies, the radio continuum spectrum and the geometry of the line
emitting region.

For modelling the RRL emission from the Circinus galaxy, we used the observed
line strength (3.2\,mJy), line width (260\,km\,s$^{-1}$), size of
the line-emitting region (250\,pc), continuum emission (225\,mJy), and
spectral index (-0.65) to constrain conditions in the ionized gas.

We used the collection of H~II regions model from Anantharamaiah et al.
(1993).  We considered a grid of models with electron temperature,
$T_{\mathrm{e}}$, in the range 1000\,K to 12\,500\,K,
electron density, $n_{\mathrm{e}}$, in the range 10\,cm$^{-3}$ to $10^{6}$\,cm$^{-3}$,
and effective diameter of the line-emitting gas ($\sim$volume$^{1/3}$), $l$, 
in the range 0.01 pc to 100 pc.  

We found that $3\times10^{3}~M_{\odot}$ to $1\times10^{6}~M_{\odot}$ of
ionized gas with $T_{\mathrm{e}} \sim 5000$\,K, $n_{\mathrm{e}} \sim
500$\,cm$^{-3}$ to $5 \times 10^{4}$\,cm$^{-3}$, and with a total
effective diameter of the line-emitting gas of 3\,pc to 50\,pc produced good
matches to the line and continuum emission.  We found that almost all of the
RRL emission is due to stimulated emission amplifying the non-thermal
continuum.  The fraction of stimulated emission was around 90\,\% in all
models.  Parameters derived for typical allowed models are given in Table 3.
The allowed range of values is large since we are using a single line strength
measurement to constrain multiple parameters.  Much tighter constraints would
come from a multi-transition study, as was done for Arp\,220 by Anantharamaiah
et al. (2000).

The inferred mass of ionized gas and the Lyman continuum flux required to
maintain the ionization are summarized in Table 3.  The flux is equivalent to
the Lyman continuum output of 300 to 9000 stars of type O5, which infers a
star-formation rate of 0.2\,$M_\odot$ yr$^{-1}$ to 6\,$M_\odot$ yr$^{-1}$ when averaged over the
lifetime of OB stars.

This can be compared to star formation rates derived from other indicators
following Hopkins et al. (2003).  Taking the peak 1.4\,GHz continuum surface
brightness of 445\,mJy\,beam$^{-1}$ in the $20'' \times 19''$ beam of the ATCA
at the nucleus of Circinus (Elmouttie et al. 1998b) yields a 1.4 GHz
luminosity in the central 400 pc diameter of $9.4\times10^{20}$\,W\,m$^{-2}$
and a corresponding star formation rate of 0.86\,$M_\odot$ yr$^{-1}$.  The
IRAS $60\,\mu$m and $100\,\mu$m flux densitites yield a far-infrared (FIR)
star formation rate of 0.17\,$M_\odot$ yr$^{-1}$. The H$\alpha$-based
star formation rate could not be estimated due to a lack of published
H$\alpha$ spectrophotometry.  The U-band magnitude of 12.87 (de Vaucouleurs et
al. 1991) along with the Balmer decrement from Oliva et al. (1994) yields a
U-band star formation rate of 1.63\,$M_\odot$ yr$^{-1}$.  These estimates all
agree well with the star formation rate of 0.2\,$M_\odot$ yr$^{-1}$ to 6\,$M_\odot$ yr$^{-1}$
estimated from the RRL emission.

\section{NGC 1808}

NGC 1808, at a distance of 11 Mpc, is a spiral galaxy (Sbc pec) undergoing a
starburst, with a FIR luminosity of $2\times10^{10}~L_{\odot}$.  This
galaxy exhibits a bright nucleus, a bar, several optical hotspots (Sersic \&
Pastoriza 1965), and dusty radial filaments over the central few kpc
(V\'eron-Cetty \& V\'eron 1985).  The peculiar morphology and the starburst
activity are attributed to tidal interactions with the companion NGC 1792
(Dahlem et al. 1990).  H\,I observations show a strong outflow (Koribalski et
al. 1993a), co-spatial with the dusty filaments.  There is evidence both for
and against Seyfert activity at the nucleus (Forbes, Boisson \& Ward 1992;
Kotilainen et al. 1996).  CO and H\,I imaging (Dahlem et al. 1990;
Koribalski, Dickey \& Mebold 1993b; Koribalski et al. 1996) shows a central
concentration of gas surrounded by a $\sim 25''$ rotating molecular ring,
believed to be coincident with the inner Lindblad resonance of the kpc-scale
bar.  There is also evidence for non-circular gas motion (Saikia et al. 1990).
High resolution radio continuum images (Saikia et al. 1990, Collison et al.
1994) show that the central kpc contains a population of compact sources.  NIR
continuum and line imaging at $1.8''$ (Krabbe, Sternberg \& Genzel 1994) and
$1.0''$ (Kotilainen et al. 1996) shows a number of Br$\gamma$, [Fe\,II], and
H$_2 $ knots.  The Br$\gamma$ and radio continuum knots are spatially
correlated at a resolution of $1.8''$ but are not exactly coincident at a higher
resolution of $1.0''$.  There is no correlation between the positions of the
optical hotspots and the radio/Br$\gamma$ knots, hence the former seem to
trace regions of low dust extinction rather than regions of star formation.
Most of the radio knots have a steep spectral index and are probably SNRs with
a thermal fraction not exceeding $\sim 30$\,\% at cm wavelengths.  Using
spectrophotometric models, Krabbe, Sternberg \& Genzel (1994) and Kotilainen
et al. (1996) derive an age of greater than 40\,Myr to 50\,Myr for the nuclear
starburst and much younger ages of less than 10\,Myr to 15\,Myr for the
circumnuclear hotspots.  Tacconi-Garman, Sternberg \& Eckart (1996) imaged the
galaxy in the NIR K band at $0.6''$ resolution and detected a number of
unresolved sources, believed to be young massive super star clusters.
Assuming a decaying star-formation rate, they derive ages similar to the
earlier work for the nucleus and the hotspots.  However, they also derive a
much larger age of $\sim 200$\,Myr for all of the sources assuming a history
of continuous star formation.

\subsection{Results}

The brightest continuum source seen in Fig 4 is the nucleus, and the second
brightest, labelled `hotspot' was seen previously in the 1.4\,GHz image of
Saikia et al. (1990). Based on the 1.4\,GHz and 8.3\,GHz continuum flux
densitites measured with a $3.1''$ beam, the spectral indices of the nucleus and
the hotspot are calculated to be $-$0.57 and $-$0.65 respectively.  Higher
resolution (sub-arcsecond) images show that the hotspot consists of two
compact distinct sources.  These are referred to as A10 and A11 by Collison et
al. (1994). Though diffuse Br$\gamma$ emission is seen in a region
encompassing both sources, only a single Br$\gamma$ knot is seen in this
region and it lies very close to A10 (Krabbe, Sternberg \& Genzel 1994,
Kotilainen et al. 1996).  Hence there seems to be more thermal gas towards
A10 than A11, even allowing for dust extinction. The radio continuum emission
from A10 is weaker than from A11 in the high resolution images. However,
images at a lower resolution of $3.1''$ at 8.3\,GHz (this work) and 1.4\,GHz
(Saikia et al. 1990) show that the smooth emission peaks much nearer the
position of A10, indicating the presence of substantial diffuse emission
around this source.

\begin{figure}[h]
  \centering
  \includegraphics[scale=0.4]{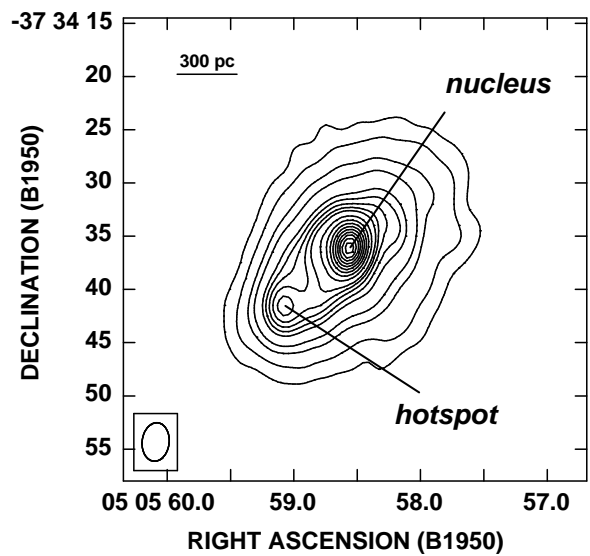}
  \caption{8.3 GHz continuum images of 
NGC 1808 with the VLA in CnB configuration. The synthesized beam is
$3.6'' \times 2.5''$ at a P.A. of $-$6$^{\circ}$. The rms noise is
35\,$\mu$Jy\,beam$^{-1}$ and the contour
levels plotted are (10, 20, 40, 60, 100, 120, 140, 160, 180, 220, 260, 
300, 340, 380, 420, 460, 500, 540)$\times$rms.}
\end{figure}

H92$\alpha$ line emission integrated over the central $18'' \times 20''$
was detected in the DnC configuration data (Fig 5), extending between
750\,km\,s$^{-1}$ and 1100\,km\,s$^{-1}$.  This velocity
range is consistent with the H\,I velocity range of (800 to
1200)\,km\,s$^{-1}$ (Koribalski et al. 1993a), considering the low signal-to-noise
ratio of the RRL detection.  The observed FWHM of the
line is (339 $\pm$ 75)\,km\,s$^{-1}$ and the deconvolved line width,
after correcting for offline Hanning smoothing and spectral resolution, is
about 210\,km\,s$^{-1}$.  The peak line emission (470
\,$\mu$Jy\,beam$^{-1}$) is only 3.9 $\sigma$.

\begin{figure}[h]
  \centering
  \includegraphics[scale=0.4]{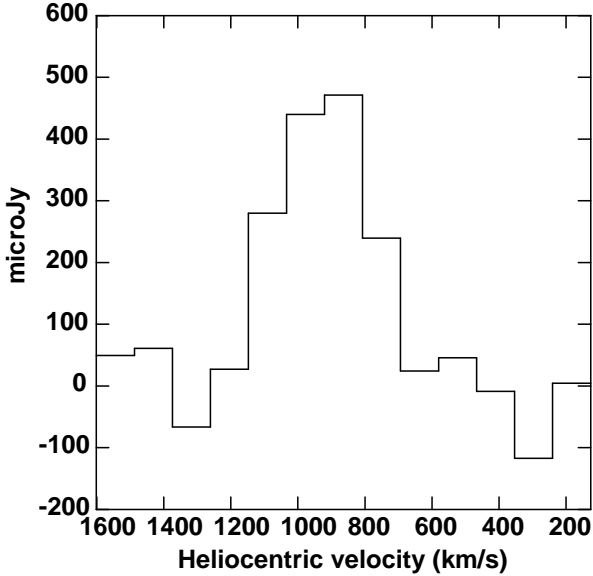}
\vspace{0.5cm}
   \caption{The Hanning smoothed H92$\alpha$ 
    spectrum in NGC 1808 for the DnC configuration data, integrated over a 
    box of size of $18'' \times 20''$ centred on RA 05 05 58.58 dec -37 34 36.5.
    The synthesized beam is $8.2'' \times 6.6''$ and
    the 1$\sigma$ noise in the channel image is 140\,$\mu$Jy\,beam$^{-1}$.
    The peak of the line integrated over the region is 0.47\,mJy
    and the continuum emission integrated over the same region is 101\,mJy
    giving a line-to-continuum ratio of 0.005.
}
\end{figure}

\subsection{Modelling the ionized gas}

We modelled the thermal gas in NGC\,1808 using the measured continuum and line
strengths at 8.3\,GHz and a spectral index of -0.7 at 8.3 GHz.  We modelled
the line-emitting gas as a collection of H~II regions.  For each combination
of input parameters in a given model and assuming conservatively that the
thermal gas was spread homogeneously over the entire region, we corrected the
intensity of the non-thermal radiation for absorption by the thermal gas and calculated
the expected free-free continuum emission from the model H~II regions.  The resulting 
free-free emission was constrained to be less than the observed thermal emission of 30\,mJy,
which was estimated from the total observed continuum emission of 101\,mJy
assuming a thermal fraction of 30\,\%.

The allowed range of electron density is 100\,cm$^{-3}$ to 1000\,cm$^{-3}$ and
the effective diameter of the line-emitting gas ($\sim$volume$^{1/3}$), $l$,
is in the range 7\,pc to 300\,pc.  The ionization rate is $3 \times
$10$^{51}$\,s$^{-1}$ to $1 \times $10$^{54}$\,s$^{-1}$.  The free-free flux
density at 8.3 GHz is $< 1.0$\,mJy to 20\,mJy which is $< 1$\,\% to 20\,\% of the
total continuum.

Model solutions were calculated for $T_{\rm e}$ of
5000\,K, 7500 K, and 10\,000\,K.  The derived parameters varied by much less than the range in the
allowed parameter values for any given $T_{\rm e}$ value.  Thus, we present
parameters for $T_{\rm e}$ = 5000\,K, but the values for $T_{\rm e}$ =
7500\,K or 10\,000\,K would be almost the same.
The fraction of stimulated emission, as for the Circinus galaxy, was around 90\,\%
in all models. 

Using the derived value of $N_{\rm{Lyc}}$= $3 \times 10^{51}$\,s$^{-1}$
to ${\bf 1
\times 10^{54}}$\,s$^{-1}$, the Br$\gamma$ flux is expected to be ${\bf 3 \times
10^{-18}}$\,W\,m$^{-2}$ to ${\bf 9 \times 10^{-16}}$\,W\,m$^{-2}$.  For comparison,
Krabbe, Sternberg \& Genzel (1994) measured the Br$\gamma$ flux from the
hotspot region to be 2.6$\times$10$^{-17}$\,W\,m$^{-2}$ and the data of
Kotilainen et al. (1996) yield 1.6$\times$10$^{-17}$\,W\,m$^{-2}$.  Both
measurements are within the broad range expected given the (extinction-free) RRL
strength.

To estimate the effect of small changes in the $b_{n}$ and $\beta$ coefficients 
on the derived physical parameters of the ionized gas, which is a concern
since the line emission is strongly affected by stimulated emission, we note that when we  
ran models for adjacent lines having roughly the same observed parameters, we 
obtained essentially the same values of derived parameters and conclude that the effects
of changes are small.

Parameters derived for typical allowed models are given in Table 3.

\section{The Non-Detections}

The H92$\alpha$ line was not detected in the other galaxies observed with the
ATCA (NGC\,6221, IC\,5063, IRAS\,18325-5926, and NGC\,7552) or with the VLA
(NGC\,4038/9, II\,Zw\,40, NGC\,7552, and VV\,114).  Upper limits were
0.8\,mJy\,beam$^{-1}$ to 1.4\,mJy\,beam$^{-1}$ ($3\sigma$) for the ATCA and
0.17\,mJy\,beam$^{-1}$ to 0.24\,mJy\,beam$^{-1}$ ($3\sigma$) for the VLA
observations (Tables 1 and 2).  The continuum images are shown in Fig 6.

{\bf IC\,5063} is classed as a NLRG, having $5\times10^{23}$\,W\,Hz$^{-1}$ at
1.4\,GHz.  The nuclear component was resolved into a $4''$-long linear triple
structure at 8.6\,GHz by Morganti et al. (1998).  Our 8.4\,GHz image (Fig 6a)
with the larger beam of $6.4'' \times 5.8''$ did not resolve the jet, as
expected.

We used the upper limits on the H91$\alpha$ + H92$\alpha$ line emission to
constrain the properties of the nuclear emission-line regions.  We
assumed that the line FWHM was the same as the velocity resolution of the
data (from Tables 1 \& 2).  We modelled each source as a collection of
uniform-density spherical H\,II regions.  The model solutions were constrained
to produce an 8.3\,GHz free-free emission equal to 30\,\% of the continuum
flux density integrated over the central region and an H92$\alpha$ line
emission equal to the observed upper limit.  The derived value for the gas
density varied between 100\,cm$^{-3}$ and $5 \times 10^{5}$\,cm$^{-3}$ and the
total effective diameter of the H\,II regions varied between 5\,pc to 200\,pc
with the effective size decreasing with increasing density.  The total flux of
ionizing photons required to maintain such nebulae on the limit of our
detectability ranged between $3 \times 10^{51}$\,s$^{-1}$ and $2 \times 10^{54}$\,s$^{-1}$
depending on density and size of the individual H\,II regions.  This provides
upper limits to the number of O5 stars powering the ionization in the range
100 to 70\,000, with the higher values corresponding to those models whose
combination of parameters yield high free-free continuum emission from the
ionized gas.

Parameters derived for typical allowed models are given in Table 3.

{\bf NGC\,6221} contains two emission-line regions in the nucleus - a
low-excitation H~II region probably photoionized by hot, young stars
and a faint high-excitation region with relatively broad emission
lines characteristic of Seyfert 2 nuclei (Veron et al. 1981).
Our continuum image at 8.3\,GHz (Fig 6b) shows emission from
the nucleus, bar and spiral arms, consistent with that seen at 1.4\,GHz by
Koribalski \& Dickey (2004).

We used the upper limits on the
H91$\alpha$ + H92$\alpha$ line emission to constrain the properties of the
nuclear emission-line regions following the approach described for IC\,5063.
Parameters derived for typical allowed models are given in Table 3.

{\bf IRAS\,18325-5926} is an ultraluminous infrared galaxy with a compact
central radio component (size $\le 0.1''$) which accounts for one fifth of the
total emission at 2.3\,GHz (Roy et al. 1994).  Our 8.3\,GHz image in Fig 6c is
consistent with that of Bransford et al. (1998), showing an unresolved
component with no significant extended structure.  The flux density in our
observation, 30\,mJy\,$\pm$\,4\,mJy, is lower than the
36.8\,mJy\,$\pm$\,0.3\,mJy measured by Bransford et al. (1998) (after
interpolating between their 4.8\,GHz and 8.6\,GHz measurements).  The source
might therefore have varied between 1993 Jul and 1995 Nov.

We used the upper limits on the
H91$\alpha$ + H92$\alpha$ line emission to constrain the properties of the
nuclear emission-line regions following the approach described for IC\,5063.
Parameters derived for typical allowed models are given in Table 3.

{\bf NGC\,4038/9} is a pair of interacting galaxies with extended tidal tails
and numerous star clusters, including super star clusters (Whitmore \&
Schweizer 1995, Whitmore et al.  1999).  Neff \& Ulvestad
(2000) imaged the multi-frequency radio continuum emission in this galaxy at
high resolution ($1.0''$ to $2.6''$) and identified a host of compact sources,
some of which exhibit a spectral index flatter than $-$0.4, indicating a high
thermal gas fraction.  Our continuum image (Fig 6d) is consistent with theirs,
showing the same collection of compact sources immersed in diffuse emission.
Our image is at lower resolution and has a factor two higher noise.

We used the upper limit on the RRL emission to constrain the properties of the
strongest thermal source {\em 54.96$-$06.1} (see Fig 6d). Neff \& Ulvestad
(2000) determined the diameter of this source to be 70\,pc ($0.7''$) and
derived an rms density of $\sim$~400\,cm$^{-3}$.

We modelled this source as a collection of uniform-density spherical H~II
regions, constrained by the H92$\alpha$ upper limit, the 8.3\,GHz flux density
of Neff \& Ulvestad (2000) as an upper limit, the observed 4.8\,GHz to
8.3\,GHz spectral index, and the infra-red excess (IRE), $L_{\rm FIR}/N_{\rm
  Lyc}$ (derived from $L_{\rm FIR}$=1.3$\times$10$^{36}$\,W; Neff \& Ulvestad
(2000) and the $N_{\rm Lyc}$ of the models) to exceed unity. No solutions were
obtained for the thermal fraction of the 8.3 GHz continuum $f_{\rm
  th}$\,$>$\,0.6 if the RRL constraint was included.  The derived values for
the gas density and the size of the H~II region varied between
5$\times$10$^2$\,cm$^{-3}$ to 10$^4$\,cm$^{-3}$ and 40\,pc to 5\,pc, and the
IRE was 6 to 10. The H92$\alpha$ line strength is $\sim$\,80\,\% of the
3$\sigma$ upper limit if the thermal fraction is not less than 30\,\% and hence
might be detectable with sensitive observations. However, the observed flat
spectral index of $-$0.26 $\pm$ 0.13 cannot be reconciled with the low derived
thermal fraction unless there is additional low density thermal gas which does
not emit 8.3 GHz RRL.  It is surprising that we do not detect RRL emission
from other regions in the galaxy. Since it is unlikely that the starburst has
declined rapidly over the last million years, a search for RRLs at much lower
resolution, possibly using the GBT, might prove more fruitful.

{\bf II\,Zw\,40} was imaged by Beck et al. (2002) at 15\,GHz using the VLA and
discovered four unresolved objects.  They identified the continuum emission as
free-free and suggested that these are similar to the young `supernebulae'
discovered in NGC\,5253 (Turner, Beck \& Ho 2000) and He\,2-10 (Kobulnicky \&
Johnson 1999).  They show that these supernebulae account for almost all of
the 12\,$\mu$m emission from the inner core.  We show our continuum image in
Fig 6e, which, as expected due to the lower resolution, shows the multiple
sources as a single unresolved component.

We used the upper limits on the H92$\alpha$ + H93$\alpha$ line emission with a
model consisting of an individual spherical H~II region of uniform density,
constrained to produce a 15\,GHz continuum free-free emission of 0.6\,mJy.
Solutions were obtained for $n_{\rm e}$\,$\geq$\,5000\,cm$^{-3}$.  The
expected 8.3 GHz H92$\alpha$ line flux for the various model solutions for all
four objects is 60\,\% to 70\,\% of the observed 3$\sigma$ upper limit for
densities between 5000\,cm$^{-3}$ and 10\,000\,cm$^{-3}$, is about 40\,\% for
$n_{\rm e}\,\sim\,$50\,000\,cm$^{-3}$ and decreases further for higher
densities.

{\bf NGC\,7552} contains three star forming rings of radii 1.0 kpc, 1.9 kpc,
and 3.4 kpc at the co-rotation and Lindblad resonance radii of the bar.  The
inner-most ring is seen in radio continuum (Forbes, Kotilainen \& Moorwood
1994).  Our continuum image (Fig 6f) shows the ring structure seen previously
by Forbes et al.  A number of Br$\gamma$ knots lie along the inner ring which
show weak correlation with radio knots, but these are not spatially resolved
in our observations.

We used the upper limits on the H91$\alpha$ + H92$\alpha$ line emission to
constrain the properties of the nuclear emission-line regions.  Parameters
derived for typical allowed models are given in Table 3.

{\bf VV\,114} is undergoing a merger and the two nuclei VV\,114\,E
and VV\,114\,W are separated by $15''$ (5.6 kpc).  The E and W components are
barely resolved in the continuum image shown in Fig 6g.

The model for the thermal component of VV 114 E was constrained by the
upper limit to the H92$\alpha$ line emission and a free-free emission of
$f_{\rm th} \times S_{\rm c}$ where $S_{\rm c}$ is 27.2\,mJy at 8.3\,GHz. The
IRE was constrained to be greater than unity by assuming that the FIR
luminosity of VV 114 E is 1$\times$10$^{11}\,L_{\odot}$.  Consistent solutions
were obtained only for a thermal fraction $\leq$\,20\,\%, i.e., $S_{\rm
  c}$\,$\leq$\,5.4 mJy.  If this thermal gas is optically thin at 8.3\,GHz,
then the ionization rate is $<\,3 \times$10$^{54}$\,s$^{-1}$.

\begin{sidewaystable}
\caption[]{Derived properties for the sample, using model results for $T_{\rm
    e} = 5000$\,K and assuming the line width was equal to the velocity resolution
 of the data }
\label{ModelRes1}
\scriptsize
\begin{center}
\begin{tabular}{llllllllll}
\hline
\hline
\noalign{\smallskip}
                            & Circinus galaxy     
                            & NGC 1808
                            & NGC 6221  
                            & NGC 7552  
                            & IC 5063 \\
\noalign{\smallskip}
\hline
\noalign{\smallskip}
Electron temperature        & 5000\,K
                            & 5000\,K
                            & 5000\,K
                            & 5000\,K
                            & 5000\,K \\

Electron density            & 500\,cm$^{-3}$ to 50\,000\,cm$^{-3}$
                            & 100\,cm$^{-3}$ to 1000\,cm$^{-3}$ 
                            & 500\,cm$^{-3}$ to 10\,000\,cm$^{-3}$
                            & 50\,cm$^{-3}$ to 1000\,cm$^{-3}$
                            & 100\,cm$^{-3}$ to $5 \times 10^{5}$\,cm$^{-3}$ \\

Effective size $^a$         & 3\,pc to 50\,pc
                            & 7\,pc to 300\,pc
                            & $<$ 6\,pc to 70\,pc
                            & 
                            & $<$ 5\,pc to 200\,pc \\

Total ionized gas mass      & $3\times10^{3}~M_{\odot}$ to
                              $1\times10^{6}~M_{\odot}$
                            & $4000~M_{\odot}$
                            & $ < (0.03$ to $2)\times10^{6}~M_{\odot}$
                            & 
                            & $ < (10^{4}$ to $2\times10^{7})~M_{\odot}$ \\

$N_{\mathrm{LyC}}$          & $1\times10^{52}$~s$^{-1}$ to
                              $3\times10^{53}$~s$^{-1}$
                            & $3 \times 10^{51}$~s$^{-1}$ to $1 \times
                              10^{54}$~s$^{-1}$
                            & $ < (0.6$ to $8)\times10^{53}$~s$^{-1}$
                            & $ < 2\times10^{54}$~s$^{-1}$
                            & $ < (0.003$ to $2)\times10^{54}$~s$^{-1}$ \\

No. O5 stars                & 300 to 9000
                            & 70 to $10^{4}$
                            & $<$ (2000 to 20\,000)
                            & $< 10\,000$
                            & $<$ (100 to 70\,000) \\

Thermal fraction at 5 GHz 
                            &
                            & $< 1$\,\% to 20\,\%
                            & 13\,\%
                            & $< 70$\,\% 
                            & \\

\hline
\hline
\noalign{\smallskip}
                            & IRAS 18325-5926
                            & NGC 4038/9
                            & II Zw 40 
                            & VV 114 \\
\noalign{\smallskip}
\hline
\noalign{\smallskip}
Electron temperature        & 5000 K
                            & 7500 K
                            & unconstrained
                            & unconstrained \\

Electron density            & 100\,cm$^{-3}$ to $1 \times 10^{5}$\,cm$^{-3}$
                            & 500\,cm$^{-3}$ to 10\,000\,cm$^{-3}$
                            & 5000\,cm$^{-3}$ to 50\,000\,cm$^{-3}$
                            & unconstrained \\

Effective size $^a$         & 6\,pc to 600\,pc
                            & 5\,pc to 40\,pc
                            & unconstrained
                            & unconstrained \\

Total ionized gas mass      & $ < (5 \times 10^{4}~M_{\odot}$ to $4 \times 10^{8}~M_{\odot})$
                            & $ < (2 \times 10^{4}~M_{\odot}$ to $2 \times 10^{5}~M_{\odot})$
                            & unconstrained
                            & unconstrained \\

$N_{\mathrm{LyC}}$          & $ < (0.008$ to $2)\times10^{55}$~s$^{-1}$
                            & unconstrained
                            & unconstrained
                            & $< 3 \times 10^{54}$\,s$^{-1}$ \\

No. O5 stars                & $< 2000$ to $6 \times 10^{5}$
                            & unconstrained
                            & unconstrained
                            & $< 10^5$ \\

Thermal fraction at 5 GHz   &
                            & $< 60$\,\% 
                            &
                            & $\leq 20$\,\% \\

\noalign{\smallskip}
\hline
\end{tabular}
\end{center}
$^{a}$ The total size, ie (total volume of all H\,II regions)$^{1/3}$ \\
\end{sidewaystable}

\section{Conclusions}

We have detected H91$\alpha$ or H92$\alpha$ lines in emission in the Circinus
galaxy and NGC\,1808 with integrated flux densities of 3.2\,mJy and
0.47\,mJy respectively using the ATCA and the VLA.  We
established upper limits on the RRL strength in NGC\,4038/9, II\,Zw\,40,
NGC\,6221, NGC\,7552, IRAS\,18325-5926, IC\,5063, and VV\,114.  The detected
line strengths infer ionized gas masses of 3000\,$M_{\odot}$ to $10^{6}
M_{\odot}$ in the Circinus galaxy and $10^4 M_{\odot}$ in NGC\,1808, and
corresponding star formation rates of 0.2\,$M_\odot$ yr$^{-1}$ to 6\,$M_\odot$
yr$^{-1}$ and 0.3\,$M_\odot$ yr$^{-1}$ to 0.6\,$M_\odot$ yr$^{-1}$, depending
on the model conditions.  The star formation rate estimated from the RRL
detection in Circinus agrees well with rates estimated from radio, FIR, and
U-band luminosities using previously-calibrated relations.

Since the detectability of RRL emission depends so sensitively on the gas
density, future prospects for using RRL emission as diagnostics of star
formation activity are good if one observes multiple transitions and includes
higher frequencies.  Multi-transition studies are important to provide
sensitivity to a wide range of gas densities and the details of which
particular transitions are detected provide the density distribution of the
ionized gas.  The line strengths have been predicted and found to increase
with frequency, making studies at 22\,GHz, 43\,GHz, and 86\,GHz attractive due
to their larger line-to-continuum ratios.  Further, the interpretation of line
strengths at higher frequency is simplified since stimulated emission is much
less important towards high frequency, giving a direct relationship
between RRL emission strength and the ionized gas mass, and hence to the star
formation rate.

\begin{acknowledgements}

The National Radio Astronomy Observatory is a facility of the National
Science Foundation operated under cooperative agreement by Associated
Universities, Inc.  The Australia Telescope Compact Array is part of the
Australia Telescope, which is funded by the Commonwealth of Australia for
operation as a National Facility managed by CSIRO.

\end{acknowledgements}

\onecolumn

\begin{figure}[h]
  \centering
  \includegraphics[scale=0.5]{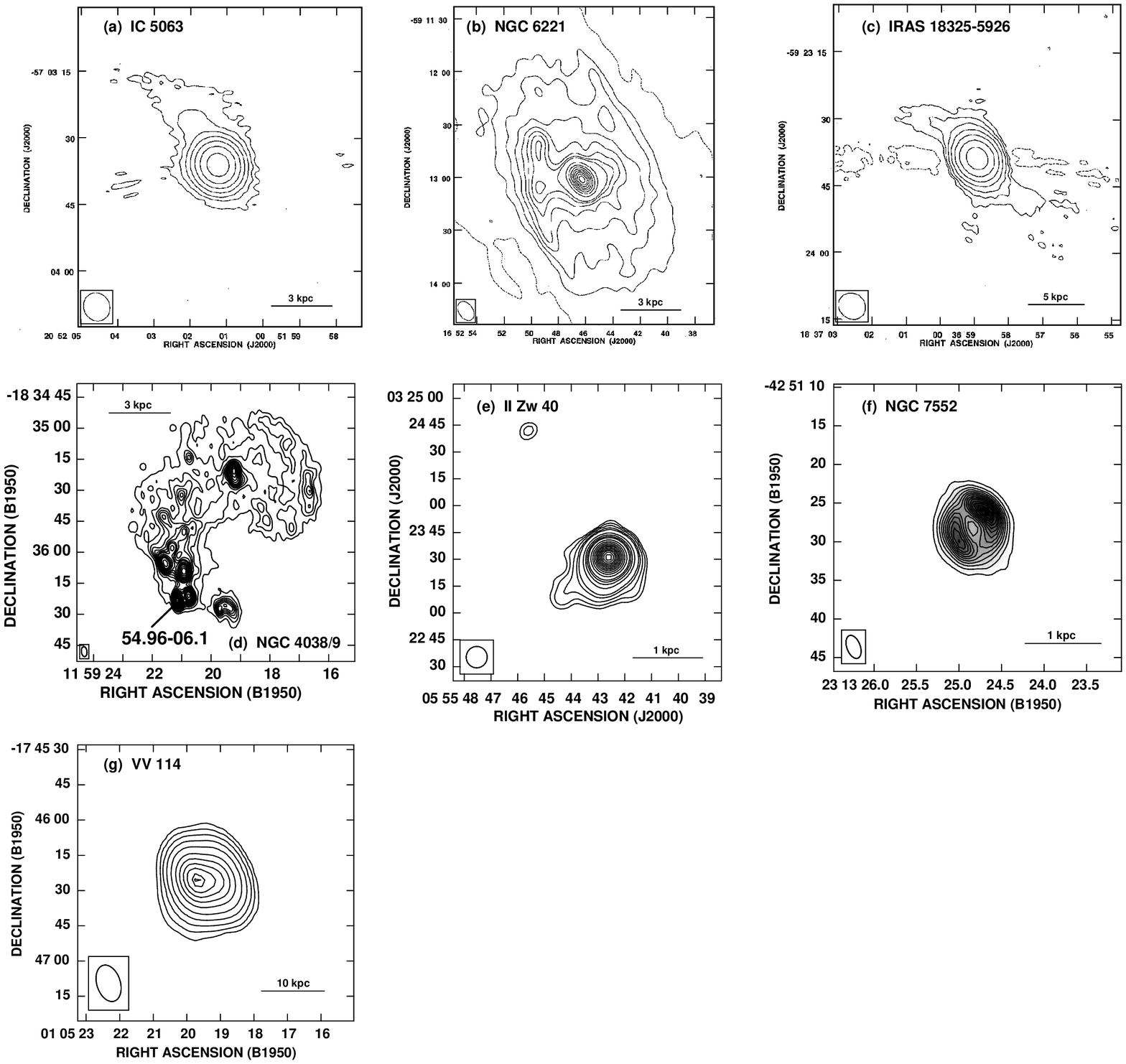}

  \caption
  {{\bf(a)} 8.4 GHz continuum image of IC 5063 using the ATCA.
    The synthesized beam is $6.4'' \times 5.8''$ at a P.A. of
    24$^{\circ}$.  The peak brightness is 180\,mJy\,beam$^{-1}$ and 
    the contour levels are at -1.8, 1.8, 3.6, 7.2, 14.4, 28.8 and 
    57.5\,mJy\,beam$^{-1}$.
  {\bf(b)} 8.4 GHz continuum image of NGC 6221 using the ATCA.
    The synthesized beam is $11.8'' \times 7.9''$ at a P.A. of
    32$^{\circ}$.  The peak brightness is 20.2\,mJy\,beam$^{-1}$ and 
    the contour levels are at -0.4, 0.4, 0.8, 1.2, 1.6, 2.0, 2.4, 2.8, 
    4.8, 6.9, 8.9, 10.9, 14.9, 14.9, 17.0 and 19.0\,mJy\,beam$^{-1}$.
  {\bf(c)} 8.3 GHz continuum image of IRAS 18325-5926 using the ATCA.
    The synthesized beam is $6.3'' \times 5.6''$ at a P.A. of
    50$^{\circ}$.  The peak brightness is 30.9\,mJy\,beam$^{-1}$ and 
    the contour levels are at -0.3, 0.3, 0.6, 1.2, 2.5, 4.9, 9.9 and 
    19.8\,mJy\,beam$^{-1}$.
  {\bf(d)} 8.3 GHz VLA continuum image of NGC 4038/9 using the C configuration.  The
    rms noise in the image is 25 $\mu$Jy\,beam$^{-1}$ and the contour 
    levels are (5,
    10, 15, 20, 25, 30, 35, 40, 50, 60, 70, 80, 100, 120, 140, 160, 180,
    200)$\times$rms in image. The synthesized beam is
    $4.8'' \times 2.7''$ at a P.A. of 7$^{\circ}$.
  {\bf(e)} 8.3 GHz continuum image of II Zw 40 using the VLA in the D configuration.
    The synthesized beam is $11.9'' \times 11.7''$ at a P.A. of
    $-$11$^{\circ}$ and the rms in the image is 20\,$\mu$Jy\,beam$^{-1}$. 
    The contour
    levels are (5, 7, 10, 15, 20, 40, 60, 80, 100, 150, 200, 250, 300, 350,
    400, 450, 500, 550, 600, 650) times the rms.
   {\bf(f)} 8.3 GHz continuum image of NGC\,7552 using the VLA in the CnB 
    configuration
    with uniform weighting. The rms in the image is 58\,$\mu$Jy\,beam$^{-1}$ 
    and the
    contour levels are (10, 20, 30, 40, 50, 60, 70, 80, 90, 100, 110, 120,
    130)$\times$rms.  The synthesized beam is
    $3.1'' \times 1.7''$ at a P.A. of 20$^{\circ}$.  
   {\bf(g)} 8.3 GHz continuum image of VV 114 in the D configuration.  The contour
    levels are (10, 20, 40, 80, 160, 320, 500, 700, 900, 1100, 1170)$\times$
    rms in the image, which is 23\,$\mu$Jy\,beam$^{-1}$. The synthesized 
    beam is $16.0'' \times 10.0''$ at a P.A. of 18$^{\circ}$.
}
\end{figure}

\twocolumn

\end{document}